\documentclass[aps,prl,twocolumn,showpacs,preprintnumbers,amsmath,floatfix]{revtex4-1}

\usepackage{graphicx}
\usepackage{epsfig}
\usepackage{dcolumn}
\usepackage{bbm}
\begin{document}
\title{The Origin of Hadron Masses}
\author{Ying Chen}
\email{cheny@ihep.ac.cn}
\affiliation{%
{Institute of High Energy Physics, Chinese Academy of Sciences, Beijing 100049, P.R. China\\
School of Physics, University of Chinese Academy
of Sciences, Beijing 100049, P.R. China}}
\date{\today}
\begin{abstract}
The color confinement can be decently explained by assuming the global $SU(3)$ color symmetry. A hadron is viewed as a bag of a finite size, whose energy is contributed by the color fields within the hadron. In the large momentum frame, the mass of a hadron can be quantized and can be expressed as the sum of the current masses of quarks involved plus a flavor independent term which depends on a universal parameter and can be identified to be the contribution from the QCD trace anomaly. This term depicts the Regge's trajectories of light hadron masses, from which the parameter can be determined and gives the minimal vector meson mass and baryon mass to be 761 MeV and 935 MeV, respectively. The mass formula can explain many features of the hadron spectroscopy. 
\\
\end{abstract}
\maketitle
The origin of hadron masses has been a hot topic in particle physics for a long term. The key problem is that, even though the current quark masses of up ($u$) and down ($d$) quarks are a few MeV, the typical mass of a hadron made up of $u,d$ quarks, such as the proton, is 
around 1 GeV and higher (this argument does not apply to the lightest pseudoscalars, which are quasi-Golstone particles due to the spontaneous breaking of the chiral symmetry). There has not been an {\it ab initio} solution to this question ever since the emergence of the quantum chromodynamics (QCD), which is a $SU(3)$ gauge theory. If one looks into the Lagrangian of QCD, one can find that the global $SU(3)$ color symmetry is still there, even though it is gauged to give QCD. It is found that the corresponding N\"{o}ther current can be governed by the matter current of a color charge, which is conserved and gauge dependent if the field $A_\mu$ generated by the charge satisfies a specific condition~\cite{Chen:2020}. Consequently the color confinement can be derived, the minimal size and thereby the minimal energy of a color singlet system, say, a hadron, can be obtained. However, we know that hadron masses are quantized, so it is desirable to investigate the possible quantization mechanism in this framework, which is surely the major mission of this work.    

In the presence of a color matter current $j^{a_0,\mu}_M$, the classical equation of motion of QCD with the solution satisfying the constraint $A_\mu^a=V_\mu\delta_{a,a_0}$ is simplified as $\partial_\mu f^{\mu\nu}=-gj^{a_0,\nu}_M$, where and $t^{a_0}$ is the generator of $SU(3)$. Accordingly, one has the field strength $f_{\mu\nu}=\partial_\mu V_\nu -\partial_\nu V_\mu$. If we split its sulotion to $f_{\mu\nu}=f_{1,\mu\nu}+f_{0,\mu\nu}$ with $f_{1,\mu\nu}$ is a special solution of the above equation, then $f_{0,\mu\nu}$ satisfies $\partial_\mu f^{0,\mu\nu}=0$ and can be expressed as~\cite{Chen:2020}
\begin{equation}
f_{0,\mu\nu}\propto g\sigma (n_{1,\mu} n_{2,\nu}-n_{1,\nu}n_{2,\mu}),
\end{equation} 
where $n_{1}$ and $n_2$ are any constant unit vectors and the constant coefficient $g\sigma$ appears by convention. If $n_1$ and $n_2$ are chosen to be light-like, $n_1^2=n_2^2 =0$ and satisfy $n_1\cdot n_2=1$, then we have $f_{0,\mu\nu}f^{\mu\nu}_0\propto -2g^2 \sigma^2$. Note that $F_{\mu\nu}=f_{\mu\nu}\sum\limits_a t^a$ is also a solution to the equation of motion of the vacuum field $[D_\mu,F^{\mu\nu}]=0$, and that the gluon condensate $\langle tr F^2\rangle$
is nonzero, one can see that $\sigma$ has a close connection with the gluon condensate and must be nonzero. Since $f_{0,\mu\nu}$ is uniform in space and nonzero, the free energy of a free color charge is infinite such that color charge can only exist in a color singlet hadron. Thus a hadron is a spatial domain of a minimal size where the color field is confined and gives the minimal energy of the hadron. Actually, the electron energy level $E_n=-\frac{Z^2 m_e\alpha^2}{2n^2}$ of a hydrogen-like atom can be also viewed as the field energy $E_V$ in the domain enveloped by the trajectory of the electron
$E_V=\int_0^{r_n} 4\pi r^2 dr \frac{1}{2} |\mathbf{E}(r)|^2=-\frac{Z^2\alpha}{2r_n}+C$, where $Z$ is the electric charge of the nucleus, $m_e$ is the electron mass, $\alpha=\frac{e^2}{4\pi}$ is the fine structure constant,  $r_n=\frac{n^2}{Z\alpha m_e}$ is the most probable radius, and $C$ is to regularize the $r\to 0$ behavior. If there are $Z$ electrons in $n$-th energy level and their interaction with each other is ignored, then the system is electric neutral and the field flux are closed within the atom. As such the energy is $ZE_n=E_V$.
  
In analogy with this, we now consider a color singlet system of color charges. If we consider a color charge of representation $R$, then the rest part of the system 
acts as a color charge in the $R^*$ representation due to the global color $SU(3)$ symmetry. If we normalize the uniform part of the color field of $R^*$ charge as~\cite{Chen:2020}
\begin{equation}
f_{0,\mu\nu}=\sqrt{C_2(R)}g\sigma (n_{1,\mu} n_{2,\nu}-n_{1,\nu}n_{2,\mu})
\end{equation}
and choose the $n_{1,2}$ to be $n_1=(1,\mathbf{e}_1)$ and $n_2=(1,\mathbf{e}_2)$, where $\mathbf{e}_1$ and $\mathbf{e}_2$ are arbitrary spatrial unit vectors in the rest frame of the system and satisfy $\mathbf{e}_1\perp \mathbf{e}_2$, then the corresponding chromoelectric and chromomagnetic field strengths are
\begin{equation}
\mathbf{E}_0=\sqrt{C_2(R)}g\sigma(\mathbf{e}_1-\mathbf{e}_2),~~~ \mathbf{B}_0=\sqrt{C_2(R)}g\sigma(\mathbf{e}_2\times \mathbf{e}_1).
\end{equation} 
Thus we can conjecture that the energy contributed by the uniform color fields to a hadron is 
\begin{equation}
E=\frac{1}{2}(\mathbf{E}_0^2+\mathbf{B}_0^2)\sum\limits_{i=1}^K V_i
\end{equation}  
where $K$ is the number of color charges, say, $K=2$ for $q\bar{q}$ mesons and $K=3$ for $qqq$ baryons, and $V_i$ is the spatial volume enveloped by the trajectory of the $i$-th (anti)quark. In this sense, we can obtain quantized energies $E$ if $V_i$ is quantized. 

The quantization can be discussed in the large momentum frame. If a hadron is flying along the $z$ direction with a velocity $v\approx 1$, the time within it is almost frozen, the energy change between quarks can be ignored and the hadron is almost stable without decay. We can choose $\mathbf{e}_z=\mathbf{e}_1\times\mathbf{e}_2$, thus in the observer's frame, the uniform field within the hadron is
\begin{equation}\label{boost}
\mathbf{E}_0'=\gamma \mathbf{E}_0, \mathbf{B}'_0=\mathbf{B}_0-v\gamma \mathbf{e}_z\times \mathbf{E}_0,
\end{equation}  
where $\gamma =(\sqrt{1-v^2})^{-1/2}$ is the Lorentz boost factor. On the other hand, a hadron flying with a large velocity looks like a thin disc, such that the energy of the uniform field can be expressed explicitly as 

\begin{figure}[t!]
	\includegraphics[height=3.0cm]{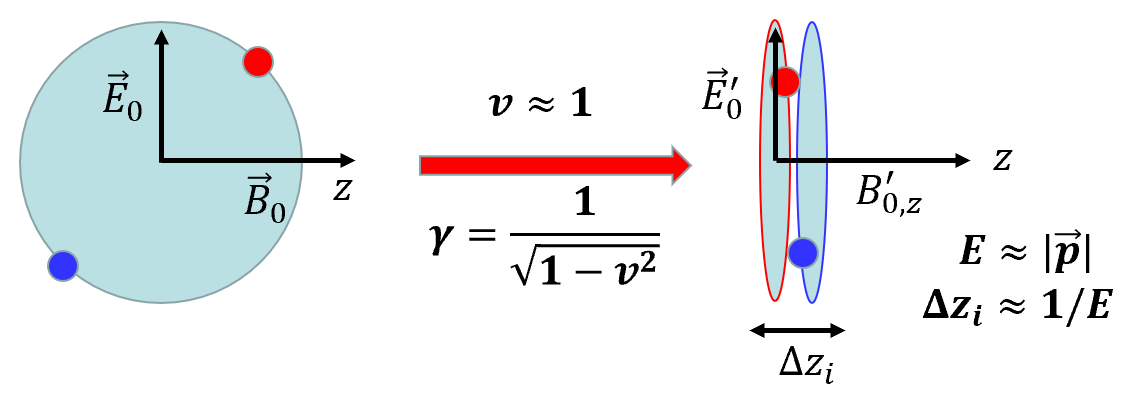}
	\caption{\label{fig:LMF} Schematic digram for a flying hadron with the velocity $v\approx 1$ along the $z$ direction. Then a spherical hadron (the left part) in the rest frame looks like a thin disk perpendicular to $z$-axis with a thickness $\Delta z_i\sim 1/E$ where $E\approx |\mathbf{p}|$ is the energy of the flying hadon. The unform fields $(\mathbf{E}_0, \mathbf{B}_0)$ are boosted to $\mathbf{E}_0',\mathbf{B}_0'$ according to Eq.~(\ref{boost}).  
	}
\end{figure}
\begin{equation}\label{energy1}
E\approx 2g^2\sigma^2 C_2(F)\gamma^2 \sum\limits_i S_i \Delta z_i,
\end{equation}
where $C_2(F)=4/3$ is the eigenvalue of the Casimir operator of the fundamental representation, $S_i$ is the perpendicular area of the domain swept by $i$-th constituent and $\Delta z_i$ is the thickness of the domain along the $z$-axis. Since the momentum carried by the $i$-th constituent is $p_{i,z}=x_i p$ with $0<x_i<1$ being the momentum
fraction of the $i$-th constituent, the uncertainty principles $\Delta z_i \Delta p_{i,z}\sim 1$ gives $\Delta z_i\approx 1/p\approx 1/E$. From Eq.~(\ref{energy1}), we now have the relation
\begin{equation}\label{energysqr}
E^2\approx  g^2\sigma C_2(F) \gamma^2 \sum\limits_i S_i.
\end{equation}

 The squared spatial momentum $\mathbf{p}_i$ of each (anti)quark can be decomposed as $\mathbf{p}_i^2 = p_{i,z}^2 + p_{i,\mathrm{kin},\perp}^2$, where $p_{i,\mathrm{kin},\perp}$ is the transverse kinetic momentum,
which, in the presence of the color field, can be expressed as
\begin{equation}
p_{i,\mathrm{kin},\perp}^2=(p_{i,\perp}+g\sqrt{C_2(F)}\mathbf{A_\perp})^2,
\end{equation}
with $p_{i,\perp}$ being the transverse canonical momentum, and the vector
potential is defined as $\mathbf{A}_\perp=\frac{1}{2}\mathbf{B}_0'\times \mathbf{r}_\perp$. In the canonical
quantization scheme, one has $\hat{p}_{i,\perp}=-i\partial_\perp$, such that $\hat{p}_{i,\mathrm{kin},\perp}^2$
satisfies the eigen-equation,
\begin{equation}
\hat{p}_{i,\mathrm{kin},\perp}^2\phi_N=\lambda_N \phi_N\equiv [(2N+1)g^2\sigma C_2(F)]\phi_N,
\end{equation}
where $\phi_N$ are eigenfunctions and the the eigenvalues are nothing but the Landau levels. In a semi-classical picture, quarks are moving in a circular trajectories
perpendicular to the $z$-axis with a semi-classical radii $\langle r_{i,\perp}\rangle
=\sqrt{\frac{2N_i+1}{g^2\sigma C_2(F)}}$. This implies that a hadron has a minimal radius in the transverse directions relative to $z$-axis, say, $r_{\perp,\rm min}\sim \sqrt{1/(g^2\sigma C_2(F))}$ (One can tentatively assume the semi-classic trajectory of the constituents are quasi-concentric such that the spatial volume of a hadron is the smallest, and thereby gives the lowest energy.)
If we take $S_i=\pi \langle r_\perp\rangle^2=\frac{\pi(2N_i+1)}{g^2\sigma C_2(F)}$,
from Eq.~(\ref{energysqr}) we have
\begin{equation}\label{eq:energy_squared}
E=\gamma \sqrt{2\pi\sigma\sum\limits_i(2N_i+1)}\equiv \gamma M_0.
\end{equation}

In the large momentum frame, the (anti)quarks are asymptotically free from each other, such that energy of each (anti)quark is 
\begin{equation}
E_q^{(i)}=\tilde{m}_i\left(1-\frac{(\hat{p}_{i,\rm{kin},\perp}^2+x^2 \mathbf{P}^2)}{x^2 E^2}\right)^{-1/2} \approx \gamma \tilde{m}_i
\end{equation} 
where $|\mathbf{P}|\sim E \to \infty$, $x$ is the momentum (energy) fraction of the constituent,  and $\tilde{m}_i$ is somewhat the current 
quark mass and depends on an intrinsic energy scale. Thus the energy of a hadron in the large momentum frame can be written as 
\begin{equation}
E=\gamma\left(\sum\limits_i \tilde{m}_i + M_0 \right)\equiv \gamma M,
\end{equation}
with $M=\sum\limits_i \tilde{m}_i +M_0$ being the rest mass of the hadron. From the point of view of the field theory, the mass of a hadron can be decomposed into $M=\langle H_m\rangle+\langle H_a\rangle$, where $H_m$ and $H_a$ are the quark mass part and the QCD trace anomaly part of the QCD Hamiltonian and $\langle\ldots\rangle$ means their expectation values between a hadron state~\cite{Ji,Yang:2014xsa,Yang:2018}. Thus $M_0$ can be identified to be the contribution of the QCD trace anomaly to the hadron mass. Note that the mass formula of $M$ is Lorentz invariant even though it is derived in the large momentum frame. Therefore, hadrons have miminal masses even in the chiral limit $\tilde{m}_i\to 0$, namely, the lowest
masses of the baryon and the meson are $M(qqq)=\sqrt{4\pi \sigma}$ and $M(q\bar{q})\sqrt{6\pi \sigma}$, respectively. Their ratio $M(qqq)/M(q\bar{q})=\sqrt{3/2}\approx 1.22$ is almost the same as the ratio of the masses of the proton and the $\rho$ meson, $M_p/M_\rho \approx 939/775\approx 1.21$. It should be stressed that the light pseudoscalars $(\pi,K,\eta)$ are Goldstone bosons owing to the spontaneous chiral symmetry breaking and are out of the scope of the discussions here. What follows is the discussion on the meson and the baryon spectra in some details.
 
{\it Meson spectroscopy:} For a quark-antiquark meson $q_1 \bar{q}_2$, since its transverse momentum is zero, it is natural to choose $N_1=N_2=N$ in Eq.~(\ref{eq:energy_squared}). Actually the Laudau levels are highly degenerate,
\begin{equation}
\lambda_n=(2N+1)C_2(F)g^2\sigma, ~~~2N=2n_r-m+|m|
\end{equation} 
where $n_r$ is the radial quantum number and $m$ is relevant to the angular momentum due to the orbital motion of the (anti)quark. For example, if $n_r=N$, then $m$ takes the values $m=0,1,2,\ldots$, while $|m|=N-n_r$ for $n_r<N$ and $m<0$. Since the (anti)quark is constrained to move in a area $A=\pi\langle r\rangle^2=\pi(2N+1)/(g^2C_2(F)\sigma)$, the degeneracy of the level is $f=\lceil Ag^2 C_2(F)\sigma/(2\pi)\rceil=N+1$, such that $|m|=0,1,\ldots,N$.  Therefore, we tentatively set the orbital angular momentum of a meson is $L=|m|$, and the meson mass can be expressed as 
\begin{equation}\label{meson}
M=\tilde{m}_1+\tilde{m}_2+\sqrt{2\pi\sigma (4N+2)}\equiv \tilde{m}_1+\tilde{m}_2+M_0
\end{equation}   
with $N=n_r+L$. It is noted that the similar formula was suggested in the phenomenological string picture~\cite{Afonin1, Afonin2}.

This meson mass formula has abundant implications: First, for a given $N$, the mass of a hadron is linear in masses of (anti)quarks involved, therefore even in the chiral limits $\tilde{m}_i\to 0$, a hadron has a nonzero mass $M_0$ which is independent of quark flavors. The linear dependence of meson masses on the current quark masses is supported by $N_f=2+1$ lattice QCD calculations~\cite{Yang:2014sea,Yang:2014xsa}. Secondly, for the hadrons made up of $u,d,s$ (anti)quarks, it is well known that they obey the flavor $SU(3)$ symmetry if taking the approximation $m_u\approx m_d \approx m_s$. Practically the $SU(3)$ flavor symmetry breaks down due to $m_u\approx m_d\ll m_s$,  
thus the meson nonet for a given $J^P$ decomposes into several isospin ($I$) mutiplets $(a,K,f_8)\oplus f_1$ where the $a(I=1)$, two $K(I=1/2)$'s and $f_8(I=0)$ are from the flavor octet and $f_1(I=0)$ is the flavor singlet. The physical $I=0$ states $f'$ and $f$ are usually the mass eigenstates mixed by $f_1$ and $f_8$ with a mxing angle $\theta$. There masses and the mixing angle $\theta$ are related by the famous Gell-Mann-Okubo relation
\begin{equation}\label{GMOR1}
\tan^2 \theta =\frac{4M_K-M_a-3M_{f'}}{-4M_K+M_a+3M_f}.
\end{equation}
where $M_X$ can be either the the mass or the mass squared of $X$ meson. 
According to Eq.~(\ref{meson}), if we take $\tilde{m}_1=\tilde{m}_2=\tilde{m_s}$ for $f'$
and $\tilde{m}_1=\tilde{m}_2=\tilde{m}_{u,d}$ for $f$, then Eq.~(\ref{GMOR1}) gives $\tan^2 \theta\approx 1/2$ and $\theta\approx 35.3^\circ$ when $\tilde{m}_{u,d}\ll \tilde{m}_s\ll M_0$ and $(\sum_i \tilde{m}_i)^2\ll M_0^2$, which is exactly what the 'ideal mixing' requires. This is supported to some extent by experimental results, for example, the mixing angles of $1^{--}(\phi(1020))$, $2^{++}(a_2(1320))$ and $3^{--}(\rho_3(1690))$ nonet
are roughly $\theta_{\mathrm{lin}}(\theta_{\mathrm{quad}})=36.5^\circ(39.2^\circ)$, $28.0^\circ(29.6^\circ)$ and $30.8^\circ(31.8^\circ)$ ($\theta_{\mathrm{lin}}$ from the mass relation and $\theta_{\mathrm{qrad}}$ from the mass squared relation of Eq.~(\ref{GMOR1})), respectively~\cite{PDG2020}. Other meson nonets (except for the ground state pseudoscalar nonet) also show this feature. In other word, the physical isoscalar mesons are dominated by either the $s\bar{s}$ or $u\bar{u}+d\bar{d}$ component. The $0^+$ states around 1.5 GeV are more interesting. There are three isocalars here, namely, $f_0(1370)$, $f_0(1500)$ and $f_0(1710)$, along with the isovectors $a_0(1450)$ and $I=1/2$ states $K^*(1430)$. If $f_0(1370)$ and $f_0(1500)$ are taken to be the $f$ and $f'$ states in Eq.~(\ref{GMOR1}), then one gets $\theta\approx 50^\circ$, which is not far from the ideal mixing and implies the 94\% $s\bar{s}$ component of $f_0(1500)$. If $f_0(1500)$ is replaced by $f_0(1710)$, then one get $\theta\approx 67^\circ$ which deviate far from the ideal mixing (As shown below, the masses of this nonet predicted by this work are $(M_a, M_K, M_f,M_{f'})\approx (1310,1405,1500,1310)\mathrm{MeV}$). In this sense, $f_0(1710)$ could be outside of the scalar nonet and could be a candidate for a scalar glueball. Of course the three isoscalars can mix to some extent, but this is an another story. 
\begin{table*}[t]
	\centering \caption{\label{tab:meson-spec}The $N$ and $L$-assignment of light hadrons. The values in bold are the predictions by Eq.~(\ref{meson}) with $2\pi\sigma\approx 0.282~\mathrm{GeV}^2 $, while the meson masses are from PDG~\cite{PDG2020}.}
	\begin{ruledtabular}
		\begin{tabular}{ccccccc}
			&  $I=1(n\bar{n})$    & $I=\frac{1}{2}(n\bar{s},s\bar{n})$       & $I=0 (s\bar{s})$ & $I=0(n\bar{n})$ & $(L,S)$ & $n^{2S+1}L_J$ \\\hline
			$M(N=0)$  (MeV)	 &  \bf{761}      & \bf{856}                  & \bf{951}           & \bf{761}          &         &               \\
			&$\rho(770)$& $K^*(892)$           & $\phi(1020)$   & $\omega(782)$ & $(0,1)$ &          $1^3S_1$     \\
			&&&&&&\\ 
			$M(N=1)$ (MeV)  &  \bf{1310}     & \bf{1405}                 & \bf{1500}          & \bf{1310}         &         &               \\
			& $\pi(1300)$       & $K(1460)$      &   $\eta(1475)$        & $\eta(1295)$ & $(0,0) $ &   $2^1S_0$   \\
			& $\rho(1450)$      & $K^*(1410)$    &   $\phi(1680)$        & $\omega(1420)$ & $(0,1)$&   $2^3S_1$   \\
			& $b_1(1235)$       &                &   $h_1(1380) $        & $h_1(1170)$    & $(1,0)$&   $1^1P_1$   \\
			& $a_0(1450)$       & $K_0^*(1430)$  &   $f_0(1500)$         & $f_0(1370)$    & $(1,1)$&   $1^3P_0$   \\
			& $a_1(1260)$       &                &   $f_1(1420)$         & $f_1(1285)$    & $(1,1)$&   $1^3P_1$   \\
			& $a_2(1320)$       & $K_2^*(1430)$  &   $f_2'(1525)$        & $f_2(1270)$    & $(1,1)$&   $1^3P_2$   \\
			&&&&&&\\
			$M(N=2)$ (MeV) & \bf{1690}     & \bf{1785}           &   \bf{1880}                & \bf{1690}           &        &              \\
			& $\pi_2(1670)$     & $K_2(1770)$    &   $\eta_2(1870)$      & $\eta_2(1645)$   & $(2,0)$&   $1^1D_2$   \\
			& $\rho(1700)$      & $K^*(1680)$    &                       & $\omega(1680)$   & $(2,1)$&   $1^3D_1$   \\
			&                   & $K_2(1820)$    &                       &                  & $(2,1)$&   $1^3D_2$   \\
			& $\rho_3(1690)$    & $K_3^*(1780)$  &   $\phi_3(1850)$      & $\omega_3(1670)$ & $(2,1)$&   $1^3D_3$   \\
			&&&&&&\\
			$M(N=3)$ (MeV) & \bf{1997}     & \bf{2092}           &     \bf{2187}  & \bf{1997}             &        &              \\
			& $a_4(1970)$       & $K_4^*(2045)$    & $f_4(2300)$         & $f_4(2050)$                 & $(3,1)$&   $1^3F_4$   \\
			&&&&&&\\
			$M(N=4)$ (MeV) & \bf{2263}     & \bf{2358}           &   \bfseries{2453}                & \bf{2263}             &        &        \\
			& $\rho_5(2350)$    & $K_5^*(2380)$  &                       &                  & $(4,1)$& $1^3G_5$   \\
			&&&&&&\\      
			$M(N=5)$ (MeV) & \bf{2501}     & \bf{2596}           &   \bfseries{2691}                & \bf{2514}             &        &              \\
			& $a_6(2510)$       &                &                       & $f_6(2510)$      & $(5,1)$& $1^3H_6$     \\       
		\end{tabular}
	\end{ruledtabular}
\end{table*}

\begin{figure}[t!]
	\includegraphics[height=5.0cm]{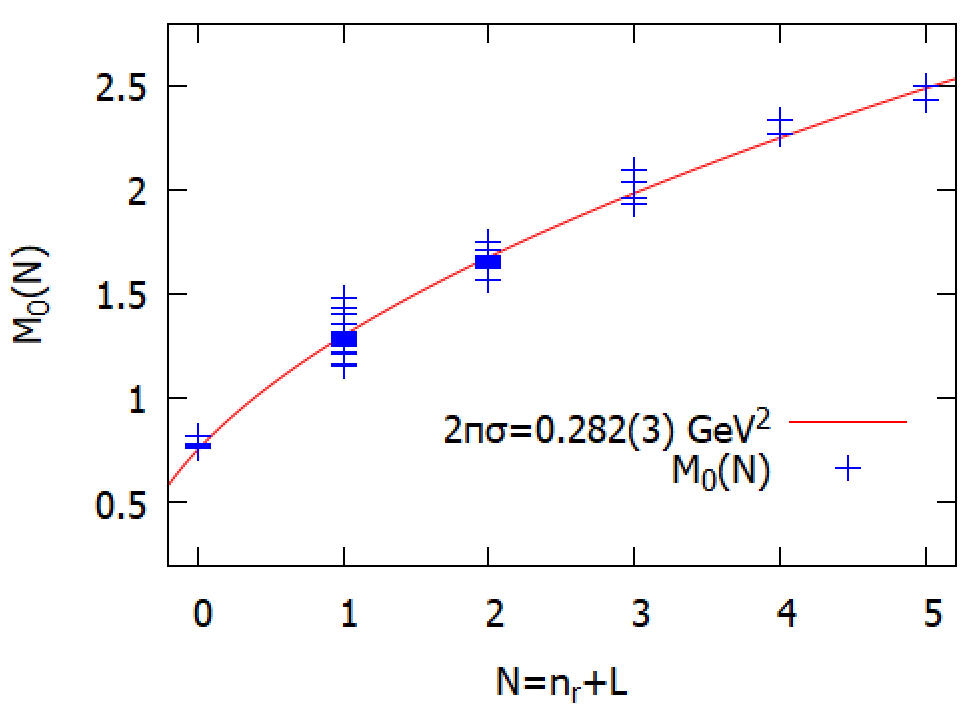}
	\caption{\label{fig:regge} The $N$-dependence of masses of light hadrons. The data points are from PDG (the experimental value is subtracted by the quark masses $\tilde{m}_1+\tilde{m}_2$, where $\tilde{m}_i\approx 5$ MeV for $u,d$ quarks and $\tilde{m}_i\approx 100$ MeV for $s$ quark). The curve is the function $M_0(N)=\sqrt{2\pi\sigma(4N+2)}$ with the fitted $2\pi\sigma=0.282(3)\,{\rm GeV}^2$.   
	}
\end{figure}

Furthermore, Eq.~(\ref{meson}) is compatible with the experimental observed $(n_r,M^2)$ and $(L,M^2)$ Regge's trajectories~\cite{regge1,regge2}, if $\tilde{m}_i\approx 0$ is assumed. We collect the light 
mesons made up of $(u,d,s)$ quarks and assign their $(n_r, L)$'s in Table~\ref{tab:meson-spec}. To be more precise, we tentatively set $\tilde{m}_u=\tilde{m}_d\approx 5$ MeV and $\tilde{m}_s\approx 100$ MeV and subtract them from meson masses according to their quark configurations, then we get the $M_0(N)$ for each meson. After that, we plot $M_0(N)$ versus $N$ in Fig.~\ref{fig:regge}, where the data points show the $M_0(N)$ with $N=n_r+L$ and the curve shows the the function form $M_0(N)=\sqrt{2\pi \sigma (4N+2)}$ with the fitted parameter $2\pi\sigma=0.282(3)\,{\rm GeV}^2$. As such the masses of $\rho$ and the nucleon $N$ are predicted to be $M_\rho \approx 761\,{\rm MeV}$ and $M_N\approx 935\,{\rm MeV}$, respectively. By the way, according to Ref~\cite{Chen:2020}, if one takes the string tension of $Q\bar{Q}$ potential is $\sigma_{Q\bar{Q}}=4g^2 \sigma /3\approx 0.2\,{\rm GeV}^2$, then the strong coupling constant $\alpha_s=\frac{g^2}{4\pi}$ can be estimate to be $\alpha_s \approx 0.27$, which is in the reasonable range.  

\begin{table*}[t]
	\centering \caption{\label{tab:1s_2s} The $1S-2S$ splittings (according to the quark model assignments) of $q\bar{q}$, $q\bar{Q}$, and $Q\bar{Q}$ mesons with $q,Q$ refering 
		to light and heavy quarks, respectively. All the data are from PDG 2020~\cite{PDG2020} except for the $B_c^*$ states, whose $1S-2S$ mass difference is given by LHCb~\cite{LHCb}.  }
	\begin{ruledtabular}
		\begin{tabular}{ccccccccccccc|c}
			& $\rho$ & $\omega$ & $K^*$ & $\phi$ & $D$ & $D_s^*$ & $\eta_c$ & $\psi$  & $B_c$ & $B_c^*$ & $\eta_b$ & $\Upsilon$ & $M_0$\\
			
			$M(2S)$ (MeV) & 1450 & 1420 & 1410 & 1680 &  2550 & 2700& 3639  & 3686 & 6872 &  * & 9999 &  10023          & 1300 \\                   
			$M(1S)$ (MeV) & 775  & 782  & 892  & 1020 &  1860 & 2112& 2985  & 3097 & 6275 &  * & 9399 &  9460 & 751\\
			&&&&&&&&&&&&&\\
			$\Delta M$ (MeV) & 675& 638 & 518  & 660  &  690  & 588 & 654   & 589  & 597  &  566 & 600&  567  & 549\\
		\end{tabular}
	\end{ruledtabular}
\end{table*}

\begin{figure}[t!]
	\includegraphics[height=5.5cm]{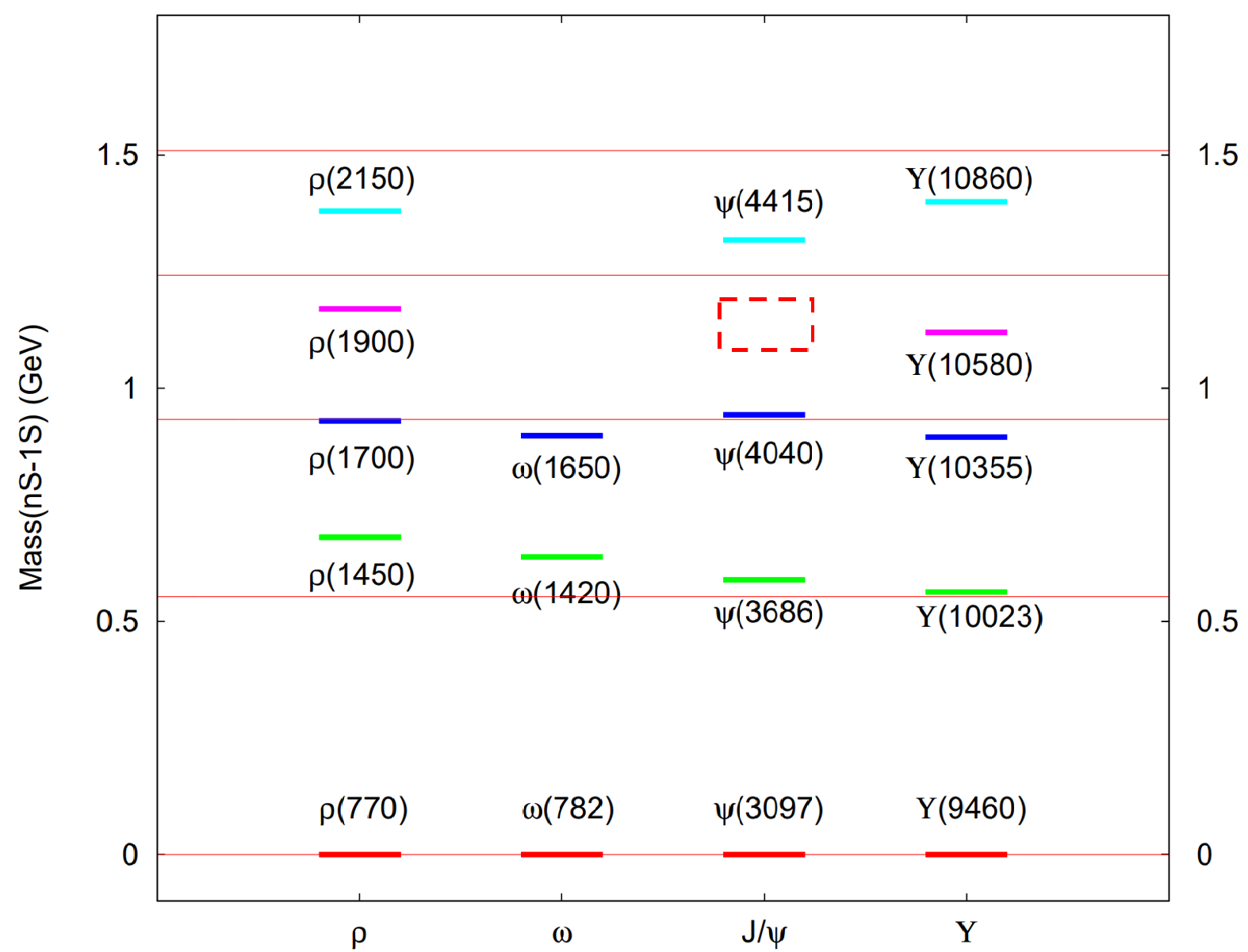}
	\caption{\label{plat}The spectrum of the $q\bar{q}=n\bar{n}$, $c\bar{c}$, $b\bar{b}$ vector mesons ($nS$ states) by setting the lowest states $\rho(770)$, $\omega(782)$, $J/\psi$, $\Upsilon$ to be on the same level. The red lines illustrate the value $\sqrt{2\pi\sigma}(\sqrt{4N+2}-\sqrt{2})$ 
		for $N=0,1,2,3,4$, respectively.
	}
\end{figure}

On the other hand, Eq.~(\ref{meson}) implies that, the mass splittings of mesons of the same quark configuration are independent of quark masses. The masses of the experimentally established $1S$ and $2S$ mesons, which correspond to $N=0$ and $N=1$ states, respectively, are collected in Table~\ref{tab:1s_2s} where most values are taken from PDG2020. except for those of $B_c^{(*)}$ states which are from LHCb results. It is interesting to note that, for all the light-light, the heavy-light and heavy-heavy mesons, the mass splittings $\Delta M=M(2S)-M(1S)$ are insensitive to quark masses and close to the value $\Delta M = \sqrt{2\pi\sigma}(\sqrt{6}-\sqrt{2})\approx 549$ MeV from Eq.~(\ref{meson}), even though the current quark masses range from a few MeV to a few GeV. The mass splittings of higher states also show the similar pattern. Figure~\ref{plat} shows the spectrum of the $q\bar{q}=n\bar{n}$, $c\bar{c}$, $b\bar{b}$ vector mesons ($nS$ states) by setting the lowest states $\rho(770)$, $\omega(782)$, $J/\psi$, $\Upsilon$ to be on the same level, where the red lines illustrate the value $\sqrt{2\pi\sigma}(\sqrt{4N+2}-\sqrt{2})$ 
for $n=0,1,2,3,4$, respectively. It is seen that the experimental values agree well with the prediction of Eq.~(\ref{meson}). However, it seems that a state (dashed line block in Fig.~(\ref{plat})) is missing between $\psi(4040)$ and $\psi(4415)$ in the $\psi(nS)$ family. $Y(4230)$ can be a candidate for this state, but its production and decay properties are very different from other $\psi$ states~\cite{Yuan:2018}. Therefore, it is intriguing that there be an additional $\psi$ state here to be discovered.  
\par
{\it Baryons spectroscopy:} Similarly, the mass formula of baryons can be written as
\begin{equation}\label{baryon-mass}
M=\sqrt{2\pi\sigma (2N+3)\sigma}+\sum\limits_{i=1}^3 \tilde{m}_i
\end{equation}
with $N=N_1+N_2+N_3$. If we take the masses of the $u,d$ quarks to be $\tilde{m}_u\approx 
\tilde{m}_d=\tilde{m}_l\approx 5$ MeV, then the masses of the lowest two baryons 
are $\sqrt{6\pi\sigma}+3\tilde{m}_l\approx 935 $ MeV ($N=0$) and $\sqrt{10\pi\sigma}+3\tilde{m}_l\approx 1202$ MeV ($N=1$), which 
are almost the masses of the nucleon and the $\Delta$ baryon. We are not yet 
very clear the interplay of the spin, the flavor and the excitation 
modes in this work. However, if we take $N$ to be even for nucleon and its excited states, we have $M(N)=(935,1420,1776,2072,2330)$ MeV for $N=0,2,4,6,8$, respectively, which comply with the $1/2^+$ nucleon spectrum $N(939),N(1440),N(1710),N(2100),N(2300)$. Similarly, if
taking $N=1,3,5,7,9$ for $\Delta$ baryons, $M(N)=(1202,1608,1930,2205,2449)$ MeV states correspond to $\Delta(1232)$, $\Delta(1600)(3/2^+)$, $\Delta(1920)(3/2^+)$, $\Delta(2300)(9/2^+)$, $\Delta(2420)(11/2^+)$. If we combine the $N$ and $\Delta$ baryons together, the experimental results show that the masses of these baryons
with different $J^P$ accumulate around the energy levels given by Eq.~(\ref{baryon-mass}).

To summarize, the global color $SU(3)$ symmetry along with the non-trivial QCD vacuum result in the color confinement. In the large momentum frame, we derive a quantized and Lorentz invariant mass formula for $q\bar{q}$ mesons and $qqq$ baryons, which is a sum of quark masses and a flavor independent term $M_0$. The term $M_0$ can be identified to be the contribution from the QCD trace anomaly and depends on a universal parameter $2\pi \sigma =0.282(3)\,{\rm GeV}^2$. This mass formula can explain may features of the hadron spectroscopy, such as the origin of the masses of $\rho$ meson and nucleon, the Regge's trajectories of light hadrons, and the almost quark mass insensitivity of many mass splittings of hadrons.

\par
 This work is supported by the National Natural Science Foundation of China (NNSFC) under Grants No.11935017, No. 11575196, No.11621131001 (CRC 110 by DFG and NNSFC) and the Strategic Priority Research Program of Chinese Academy of Sciences (No. XDB34030302). The author also acknowledges the support of the CAS Center for Excellence in Particle Physics (CCEPP).

\end{document}